\documentclass[aps,prd,superscriptaddress,twocolumn]{revtex4-1}
\usepackage{graphicx}
\usepackage[caption=false]{subfig}
\usepackage{epstopdf}
\usepackage{amsmath}
\usepackage{amsfonts} 
\usepackage{amssymb}
\usepackage{latexsym}
\usepackage{hyperref}
\usepackage[english]{babel}
\usepackage[utf8]{inputenc}
\usepackage[colorinlistoftodos]{todonotes}
\usepackage{xcolor}
\usepackage{slashed}
\usepackage{feynmp}
\usepackage{bm}
\usepackage{bbold}
\usepackage{eufrak}
\usepackage{slashed}
\usepackage{bm}  % To make greek letters in bold face {\bm \Phi}

\begin{document}
\title{A novel test of Lorentz violation in the photon sector with an LC circuit}

\author{P. C. Malta}\email{pedrocmalta@gmail.com}
\affiliation{Centro Brasileiro de Pesquisas F\'{i}sicas (CBPF), Rua Dr.~Xavier Sigaud 150, Urca, Rio de Janeiro, Brazil, CEP 22290-180}

% \ = shift + opt + 7
% { = opt + 8
% } = opt + 9

\begin{abstract}
In the presence of an external magnetic field, the Carroll-Field-Jackiw term introduces a displacement current proportional to the Lorentz-violating background that induces a time-dependent magnetic field. Axion-like particles or hidden photons could generate an analogous signal, potentially detectable with the set-up suggested by Sikivie, Tanner and Sullivan -- a sensitive magnetometer coupled to a superconducting LC circuit. We show that a similar set-up, but with an externally driven pick-up loop whose area varies harmonically at $\sim$~Hz, can be used to probe the spatial components of the Lorentz-violating background to the level of $\lesssim 10^{-31}$~GeV. This is eight orders of magnitude more sensitive than previous laboratory-based limits. 
\end{abstract}

%\pacs{11.30.Cp, 12.60.-i, 13.40.Em}
\maketitle

%%%%%%%%%%%%%%%%%%%%%%%%%%%%%%%
\section{Introduction} \label{sec_intro}
\indent

The Standard Model is an extremely successful theory firmly based on the principles of quantum mechanics and special relativity. Invariance under Lorentz transformations has been consistently tested and so far no deviations have been found~\cite{Matt}. Nonetheless, some promising extensions of the Standard Model, like string theory, allow Lorentz symmetry to be violated~\cite{Kost}. In fact, Lorentz-symmetry violation (LSV) may be introduced in all sectors of the Standard Model, which is then generalized in the so-called Standard Model Extension (SME)~\cite{Kost2, Kost3}. For experimental limits on its various sectors, see ref.~\cite{Tables} and references therein.

Carroll, Field and Jackiw (CFJ) proposed a CPT-odd, Chern-Simons-like Lagrangian in which the electromagnetic fields are coupled to the 4-vector $k_{\rm AF}$ via~\cite{CFJ}
\begin{equation}
\mathcal{L}_{\rm CFJ} = \frac{1}{2} \epsilon_{\mu\nu\alpha\beta} \left(k_{\rm AF}\right)^{\mu} \! A^{\nu}F^{\alpha\beta} \; , \label{CFJ}
\end{equation}
where $A^{\mu}=\left(\phi, \, {\bf A} \right)$ is the 4-potential and $F^{\mu\nu} = \partial^\mu A^\nu - \partial^\nu A^\mu$. Note that $\left[k_{\rm AF}\right] = $~mass. This term is gauge invariant if $k_{\rm AF}$ is non-dynamic, thus providing a preferred direction in space-time and breaking Lorentz invariance.

Maxwell's electrodynamics is modified by eq.~\eqref{CFJ} and, since it has been very well tested, the CFJ background is equally well constrained. Besides theoretical investigations, many experimental tests of the CFJ model have been proposed. As discussed already in CFJ's seminal work, the presence of this LSV term would induce a rotation of the polarization of the light from distant radio galaxies, whose non-observation led to the upper bound $k_{\rm AF}^{\rm Z} \lesssim 10^{-42} \, {\rm GeV}$~\cite{CFJ, Goldhaber}. Current bounds from CMB polarization are one order of magnitude stronger~\cite{Kost5, Mewes2}. Laboratory tests looking for LSV-induced birefringence are usually not as stringent, reaching $k_{\rm AF}^{\rm Z} \lesssim 10^{-23} \, {\rm GeV}$, mostly due to the shorter optical path legth in comparison to astrophysical sources, but nonetheless represent complementary tests of LSV in the photon sector~\cite{Yuri}.

In this paper we discuss a new laboratory-based test of the CFJ model inspired by the proposal put forward by Sikivie, Sullivan and Tanner~\cite{Sikivie_0} in the context of photons coupled to axion-like particles (ALPs) as cold dark matter candidates~\cite{Duffy, Taoso}. The ALP-photon coupling modifies the Maxwell equations and, in the presence of an external magnetic field, creates a displacement current serving as the source for an ALP-originated magnetic field. The flux of this field through a pick-up loop generates a current in a superconducting circuit that in turn induces a magnetic field in a separate coil, which could be detected by a sensitive magnetometer (e.g., a SQUID). For first results, see ref.~\cite{ADMX_SLIC}. A similar arrangement, but without an external magnetic field, was proposed by Arias {\it et al.} to search for hidden photons~\cite{Arias_0}.

Here we show that the CFJ Lagrangian~\eqref{CFJ} modifies the Maxwell equations analogously to the ALP-photon couplings discussed in ref.~\cite{Sikivie_0}, but instead of light ALPs, it is the time component of the LSV background that couples to an external magnetic field. The same process as in the ALP case would take place and the resulting magnetic field could be detected by a magnetometer. However, due the much lower frequencies involved, the set-up discussed in ref.~\cite{Sikivie_0} must be modified to improve the sensitivity to the CFJ signal.

This paper is organized as follows: in sec.~\ref{sec_CFJ} we discuss the modified Maxwell equations with the CFJ term and in sec.~\ref{sec_setup} we trace a parallel to the analysis from Sikivie, Sullivan and Tanner to the case of the CFJ model. In sec.~\ref{sec_est} we discuss modifications to their set-up to estimate attainable sensitivities. Finally, in sec.~\ref{sec_conc} we present our closing remarks. We use natural units ($c = \hbar = 1$, $\mu_0 = 1/\epsilon_0 = 4\pi $) throughout.

%%%%%%%%%%%%%%%%%%%%%%
\section{The CFJ electrodynamics} \label{sec_CFJ}
\indent

If external sources $J^\mu = \left( \rho, \, {\bf J} \right)$ are present, the inhomogeneous equations of motion become
\begin{equation} \label{eq_motion}
\partial_{\mu}F^{\mu\nu} = 4\pi J^{\nu} -2\left(k_{\rm AF}\right)_{\mu}\tilde{F}^{\mu\nu} \; .
\end{equation}
Writing eq.~\eqref{eq_motion} in terms of electric and magnetic fields via $F^{0i} = -{\bf E}_i$ and $F^{ij} = -\epsilon_{ijk}{\bf B}_k$, we have
\begin{eqnarray}
{\bm \nabla}\cdot{\bf E} & = & 4\pi \rho + 2{\bf k}_{\rm AF}\cdot{\bf B} \label{eq_coulomb} \\
-\partial_0{\bf E} + {\bm \nabla}\times{\bf B} & = & 4\pi {\bf J} + 2k_{\rm AF}^0{\bf B} -2{\bf k}_{\rm AF}\times{\bf E} \; , \label{eq_max_amp}
\end{eqnarray}
and we note that, if external electric or magnetic fields are applied, the source terms in the Coulomb and Amp\`ere-Maxwell laws acquire novel LSV contributions.

The CFJ model has a few interesting features. In momentum space, eq.~\eqref{eq_coulomb} indicates that, in a charge-free region, the electric field is not transverse. This also implies that the Poynting vector is not parallel to the wave vector, analogously to what is encountered in anisotropic materials~\cite{Kost3, Wang}. Therefore, not unlike ALPs~\cite{Tobar}, the presence of the background induces terms playing the role of a polarization or magnetization in empty space~\cite{CFJ, Bailey}. The stability, unitarity and causality of the CFJ model were extensively discussed in the literature. For instance, it can be shown that the Hamiltonian is not positive-definite for a time-like background, even if a small non-zero photon mass {\it \`a la} Proca is introduced~\cite{Adam, Klinkhamer, UFMA}. This means that, for such a background, the theory is unstable. The unitarity of the model is also not guaranteed~\cite{Adam}. Hence, in principle, only a space-like background would give rise to an acceptable field theory and experimental searches should find stringent upper limits, particularly on the time component of $k_{\rm AF}$.

In the following we focus on the Amp\`ere-Maxwell equation~\eqref{eq_max_amp} with an external magnetic field. In this scenario, the only relevant component of the background is $k_{\rm AF}^0$, which is measured in the laboratory frame. However, it is important to note that the non-dynamic nature of the CFJ 4-vector is only evident in an inertial frame, but Earth-bound experiments do not satisfy this requirement due to Earth's sidereal and orbital motions.

A convenient choice of reference frame is the so-called Sun-centered frame (SCF)~\cite{Tables, Mewes}, which is connected to the laboratory by a Lorentz transformation given by $\Lambda^{0}_{\,\,T} = 1$, $\Lambda^{0}_{\,\,I} = -{\bm \beta}^{I}$, $\Lambda^{i}_{\,\,T} = - (R\cdot{\bm \beta})^{i}$ and $\Lambda^{i}_{\,\,I} = R^{iI}$. Here $R^{iI}$ is a spatial rotation and ${\bm \beta}$ is the velocity 3-vector of the laboratory relative to the SCF. The latter is explicitly given by
\begin{eqnarray}
{\bm \beta}^X & = &  \beta_{\oplus} \, \sin(\Omega_{\oplus}T) - \mathcal{O}\left(\beta_{L}\right) \label{beta_x} \\
{\bm \beta}^Y & = &   -\beta_{\oplus} \, \cos\eta \, \cos(\Omega_{\oplus}T) + \mathcal{O}\left(\beta_{L}\right)  \label{beta_y} \\
{\bm \beta}^Z & = &   -\beta_{\oplus} \, \sin\eta \, \cos(\Omega_{\oplus}T) \label{beta_z}  \; ,
\end{eqnarray}
where $\beta_{\oplus} \approx 10^{-4}$ and $\Omega_\oplus = 2\pi/{\rm year} \approx 0.2 \;\mu$Hz are Earth's orbital velocity and frequency. Here $\eta \approx  23.4^{\circ}$ is the inclination of Earth's axis relative to the orbital plane and $\beta_{L} \lesssim 10^{-6}$ is Earth's latitude- and time-dependent sidereal velocity~\cite{Mewes}.

By using the Lorentz transformations above we may write $k_{\rm AF}^0$ in terms of the components of the background in the SCF as
\begin{equation} \label{k0_SCF}
k_{\rm AF}^0 \simeq k_{\rm AF}^{\rm T} - {\bm \beta} \cdot {\bf k}_{\rm AF}^{\rm SCF} \; ,
\end{equation}
showing that LSV signals detected in an Earth-bound experiment would present a very broad time modulation due to Earth's motion relative to the SCF~\cite{Mewes}.

\section{The Sikivie-Sullivan-Tanner set-up} \label{sec_setup}
\indent

Let us now focus on eq.~\eqref{eq_max_amp} in the presence of an external magnetic field ${\bf B}_{\rm ext}$. Assuming that the apparatus is sufficiently well shielded from external electric fields, the time component of the CFJ background induces a current density
\begin{equation} \label{CFJ_current}
{\bf J}_{\rm CFJ} = 2k_{\rm AF}^0{\bf B}_{\rm ext} \; ,
\end{equation}
whose time dependence is determined by eq.~\eqref{k0_SCF} in the case of a static external magnetic field.

In ref.~\cite{Sikivie_0} the authors consider ALPs coupled to the electromagnetic fields via $\mathcal{L}_{\rm ALP} = -g_{a\gamma} a(t) {\bf E}\cdot{\bf B}$, where $a(t)$ is the ALP field. With this extra term we obtain modified Maxwell equations that are analogous to eqs.~\eqref{eq_coulomb} and \eqref{eq_max_amp}, but with $2k_{\rm AF}^0 \rightarrow -g_{a\gamma} \dot{a}$, where $\dot{a}$ is the time derivative of the ALP field. In the following we analyse the consequences of eq.~\eqref{CFJ_current} in analogy to the ALP discussion using the Sikivie-Sullivan-Tanner set-up detailed in ref.~\cite{Sikivie_0} (see also refs.~\cite{ADMX_SLIC, Arias_0}).

For ALPs with $m_a \sim 10$~neV we have $\lambda_a \sim 20$~m, which is larger than the typical size of the experiment $\sim \mathcal{O}(10\; {\rm m})$, so that a magnetostatic regime may be assumed~\cite{APS}. In our case the frequencies involved are much lower -- of order $\Omega_{\oplus}$ -- so we are in the same regime. In this case, the current in eq.~\eqref{CFJ_current} serves as the source of a magnetic field satisfying ${\bm \nabla}\times{\bf B}_{\rm CFJ} = {\bf J}_{\rm CFJ}$. If we make ${\bf B}_{\rm ext} = B_{\rm ext} \hat{{\bf z}}$, we have
\begin{equation} \label{B_CFJ}
{\bf B}_{\rm CFJ} = k_{\rm AF}^0 B_{\rm ext} r \hat{{\bm \phi}} \; ,
\end{equation}
where $\hat{{\bm \phi}}$ is the unitary vector in the azimuthal direction and $r$ the radial distance from the symmetry axis of the magnet bore.

Let us consider a large rectangular pick-up loop with length $\ell_m$ and radius $r_m$ conveniently placed inside the bore of a solenoid and connected to a small detection coil as illustrated in fig.~\ref{fig_sketch}. The loop is traversed by a flux $\Phi_{\rm CFJ} = 2 V_m k_{\rm AF}^0 B_{\rm ext}$, where $V_m = l_m r_m^2/4$, thus inducing a current $I_{\rm CFJ} = - \Phi_{\rm CFJ}/L$. Here $L \simeq L_m + L_d + L_c$ is the inductance of the circuit and $L_m$, $L_d$ and $L_c$ are the inductances of the pick-up loop, of the detection coil and of the coaxial cable connecting the two, respectively. Note that $L_d$ is frequency dependent~\cite{Sikivie_0}.

The current $I_{\rm CFJ}$ flows into the small detection coil of radius $r_d$ with $N_d$ turns and generates a magnetic field of magnitude
\begin{equation} \label{B_d}
B_d = 4\pi \frac{N_d V_m k_{\rm AF}^0 B_{\rm ext}}{r_d L} \; .
\end{equation}
This is essentially the signal we wish to detect with the magnetometer in the Sikivie-Sullivan-Tanner set-up. In the following we discuss the sensitivity if their set-up is used, as well as modifications necessary for the detection of our particular signal.

%%%%%%%%%%%%%%%%%%%%%%%%%%%%
\section{Detection sensitivity} \label{sec_est}
\indent

The discussion above is based on the close analogy between the CFJ Lagrangian~\eqref{CFJ} and the ALP-photon coupling. The Sikivie-Sullivan-Tanner set-up would then be useful not only to search for dark matter candidates, but also to look for LSV in the photon sector of the SME. There is however one important difference between these applications: the frequency of the signal. The target in ref.~\cite{Sikivie_0} is to detect ALPs with masses $\sim$~10~neV, corresponding to frequencies $\sim 1$~MHz. In order to amplify the signal, the LC circuit is designed so that its resonance frequency $\omega_r = 2\pi \nu_r = 1/\sqrt{LC}$ approximately matches the signal frequency given by the ALP's mass, what can be accomplished by tuning the circuit's capacitance~$C$~\cite{world_sci}. In this case, the current would be enhanced by $Q$, the quality factor of the LC circuit.

The LSV signal is crucially determined by the flux of ${\bf B}_{\rm CFJ}$ through the pick-up loop. A time dependence in the flux may be introduced through three factors: the CFJ background itself, the external magnetic field and the area of the pick-up loop. Let us consider each of them separetely. As shown in eq.~\eqref{k0_SCF}, the CFJ background varies very slowly as Earth moves relative to the SCF. Let us assume an inductance $L \sim 10\; \mu$H and a capacitance $C \sim 0.1 \; \mu$F -- a typical value for commercial capacitors. With these parameters, the LC circuit resonates at $\sim$~MHz, very far from the original signal frequency $\Omega_{\oplus} \sim \mu$Hz. Therefore, if the original Sikivie-Sullivan-Tanner set-up is used and the only time dependence is that from eq.~\eqref{k0_SCF}, the signal would be too far from resonance and would be strongly suppressed.

The second factor is the magnetic field to which the pick-up loop is exposed. Instead of a constant field, alternating-current (AC) fields could be used. Unfortunately, this approach has a number of disadvantages. High-frequency AC fields -- up to $\sim$ 0.5~MHz -- with intensities of $\mathcal{O}(0.1\; {\rm T})$ can only be produced within bore volumes of a few cm$^3$~\cite{Lenox, Mazon, Connord, Rowe}. Furthermore, solenoids that produce such high frequencies and field intensities require strong currents and thick, tightly winded coils that would generally experience significant ohmic losses. Alternatively, the pick-up loop could be placed inside a superconducting, high finesse resonant cavity designed to operate in the TE mode, where high frequencies and strong magnetic fields may be more easily produced in a larger volume. Both solutions would critically suffer from the fact that strong AC magnetic fields would induce equally strong AC electric fields. These fields would interact with the wires in the pick-up loop and induce large background currents, masking the LSV signal.

\begin{figure}[t!]
\begin{minipage}[b]{1.\linewidth}
\includegraphics[width=\textwidth]{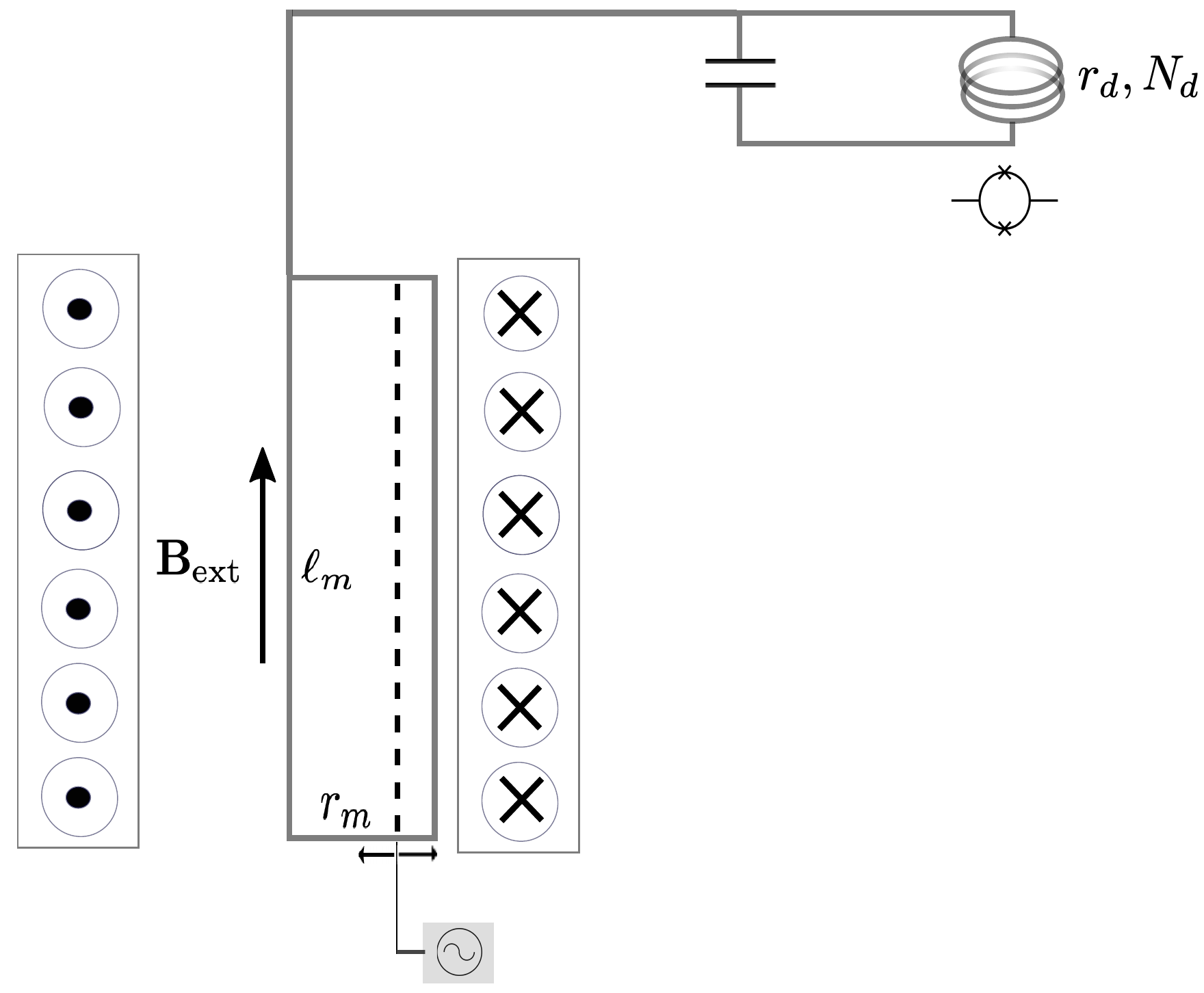}
\end{minipage} \hfill
\caption{Sketch of the Sikivie-Sullivan-Tanner set-up modified to include an externally driven wire closing the pick-up loop (dashed line) at frequencies $\nu_{\rm act} \approx \nu_r \sim$~Hz. The actuator is indicated by the gray box below the solenoid and the large capacitor is displayed near the detection coil, which is connected to the pick-up loop by a coaxial cable.}
\label{fig_sketch}
\end{figure}

Finally, the flux depends on the area through which the LSV-induced field ${\bf B}_{\rm CFJ}$ flows. Keeping the external magnetic field static and ignoring the broad time modulation due to the CFJ background, we may induce an AC LSV current by mechanically varying the area of the pick-up loop along its shortest side, $r_m$, cf.~fig.~\ref{fig_sketch},. This can be achieved by state-of-the-art actuators used, for exemple, in modern optical lithography applications, where wafer stages must be repetitively positioned with sub-$\mu$m precision within ranges of a few cm, thereby reaching accelerations of up to~$12 \; g$ -- for a review, see ref.~\cite{Schmidt}. For a harmonic movement, the acceleration $a$ and the maximal displacement $x_{\rm max}$ are connected to the driving frequency $\nu_{\rm act}$ via $a = (2\pi \nu_{\rm act})^2 x_{\rm max}$, so that $\nu_{\rm act} \simeq 2$~Hz can be achieved for $a = 10 \; {\rm m/s^2}$ and $x_{\rm max} = 5$~cm.

With the strategy outlined above it is possible to raise the signal frequency from $\mu$Hz to a few Hz. In order to gain the enhancement from the quality factor, we need to increase the capacitance of the circuit, which for the high frequencies in ref.~\cite{Sikivie_0} is of order~$\mu$F. This may be achieved by using so-called supercapacitors -- potentially several combined -- which may reach up to a few kF~\cite{Tang}. With inductances of a few ten $\mu$H, this means that the resonance frequency of the LC circuit is $\nu_r \simeq \mathcal{O}(1 \; {\rm Hz})$, which is in the range of frequencies attainable with the external actuators described above.

For the rest of the discussion we assume that the original Sikivie-Sullivan-Tanner set-up can be modified to accommodate a pick-up loop with a varying area as outlined above. The signal resonates with a frequency $\nu_r$ of a few Hz and is enhanced by the quality factor of the circuit. This adaptation is sketched in fig.~\ref{fig_sketch}, where the basic parameters of the pick-up loop and detection region are shown. The dashed line indicates the moving side of the loop and the arrows highlight the action of the actuator, which is isolated from the solenoid to avoid vibrations and thermal effects.

The placement of the pick-up loop in the magnet bore is very important. Unfortunately, just half of the diameter of the bore -- typically cylindrical -- may be used, otherwise the net flux is zero. Also, the external magnetic field must lie in the plane of the pick-up loop in order to avoid the induction of parasitic currents due to the flux of ${\bf B}_{\rm ext}$ through the time-dependent area of the tilted plane of the loop. Due to Faraday's law, ${\bf B}_{\rm CFJ}$ will induce a small AC electric field pointing in the $z$-direction that turns on the term $\sim {\bf k}_{\rm AF}\times{\bf E}$ in eq.~\eqref{eq_max_amp}. This contribution is of second order in the CFJ background and can be therefore neglected. Moreover, the CFJ contribution to the Coulomb equation~\eqref{eq_coulomb} may be ignored in comparison with the charge densities present in the system.

%due to the essentially static -- and random -- orientation of ${\bf k}_{\rm AF}$ relative to the apparatus, this term will not contribute over the measurement time, especially if the experiment runs over a long period. 

Another issue is the limitation due to the stray capacitance $\mathcal{C}$, which may cause losses for frequencies above $\omega_{\rm stray} \approx 1/\sqrt{L \cdot \mathcal{P} \mathcal{C}}$, where $\mathcal{P}$ is the length of the circuit, roughly given by $\ell_m$; cf.~fig.~\ref{fig_sketch}. The magnetostatic condition is also an important requirement that is nonetheless clearly fulfilled even for our increased frequencies. The considerations above indicate that $\nu_r$ may not exceed $\nu_{\rm stray}$, which plays the role of a cut-off frequency. For the envisaged parameters, $\mathcal{C} \simeq 15$~pF/m~\cite{Sikivie_0, Arias_0}, $L \simeq 10 \; \mu$H and $\ell_m \simeq 1-10$~m, we have $\nu_r \ll \nu_{\rm stray}$, so no cut-off applies to our analysis.

%The fulfilment of the magnetostatic condition is an important requirement. In this regime, the system is small compared with the wavelength connected with the relevant time scale~\cite{APS}. If $\ell_m$ is representative of the size of the circuit, we may associate with it the frequency $1/\ell_m$ and, by taking~$\approx 10 \; \ell_m$ as a threshold for ``large enough'', we may say that $\nu_{\rm quasi} \approx 30 \; {\rm MHz} \left( {\rm m}/\ell_m\right)$ acts as an upper limit on the frequencies satisfying the magnetostatic condition. Another issue is the limitation due to the stray capacitance $\mathcal{C}$ -- measured in F/m -- which may cause losses for frequencies above $\omega_{\rm stray} \approx 1/\sqrt{L \cdot \mathcal{P} \mathcal{C}}$, where $\mathcal{P}$ is the length of the circuit, roughly given by $\ell_m$. The considerations above indicate that $\nu_r$ may not exceed either $\nu_{\rm stray}$ or $\nu_{\rm quasi}$, which play the role of cut-off frequencies. For the envisaged parameters, $\mathcal{C} \simeq 15$~pF/m~\cite{Sikivie_0, Arias_0}, $L \simeq 10 \; \mu$H and $\ell_m \simeq 1-10$~m, we have $\nu_r \ll \nu_{\rm stray}, \nu_{\rm quasi}$, so no cut-off applies to our analysis.

%
%\begin{equation} \label{stray}
%\nu_{\rm stray} \approx 160 \; {\rm MHz} \; \sqrt{ \left( \frac{\rm m}{\mathcal{P}} \right) \!  \left( \frac{\rm pF/m}{\mathcal{C}} \right) \! \left( \frac{\rm \mu H}{L} \right) } \; .
%\end{equation}
%

Finally, let us consider the LSV signal. It is given by $I_{\rm CFJ}$ multiplied by $Q$, the quality factor of the circuit, and may be conveniently expressed as
\begin{eqnarray} \label{signal_CFJ}
I_{\rm CFJ} & = & 1.0 \times 10^{-3} \; {\rm A} \left( \frac{V_m}{{\rm cm}^3} \right) \left( \frac{\mu{\rm H}}{L} \right) \cdot  \nonumber \\
& \cdot &  \left( \frac{Q}{10^4} \right) \left( \frac{B_{\rm ext}}{\rm T} \right) \left( \frac{k_{\rm AF}^0}{10^{-23} \; {\rm GeV}} \right) \; .
\end{eqnarray}
The main noise sources were discussed in ref.~\cite{Sikivie_0}, where it is shown that the magnetometer noise is typically much lower than the thermal noise, given by $\delta I_{\rm T} = \sqrt{4 k_B T Q \Delta\nu / L \omega_r}$~\cite{Nyq}. Setting $\omega_r = 2\pi \nu_r$, the signal to noise ratio ${\rm SNR} \simeq I_{\rm CFJ} / \delta I_{\rm T}$ reads
\begin{eqnarray} \label{SNR}
{\rm SNR} & \simeq & 3.5 \times 10^{7}  \left( \frac{V_m}{{\rm cm}^3} \right) \! \left( \frac{Q}{10^4} \right)^{1/2} \! \left( \frac{\mu{\rm H}}{L} \right)^{1/2} \!  \left( \frac{B_{\rm ext}}{\rm T} \right)  \cdot  \nonumber \\
& \cdot \! & \!  \left( \frac{{\rm mK}}{T} \right)^{1/2} \! \left( \frac{\nu_r}{{\rm Hz}} \right)^{1/2} \! \left( \frac{k_{\rm AF}^0}{10^{-23} \; {\rm GeV}} \right) \; ,
\end{eqnarray}
with the bandwidth $\Delta\nu = 1$~mHz held fixed. The signal, whose frequency is now determined by the external driving actuator, is assumed to be coherent throughout the measurement time of, say, $10^3$~s~\cite{Sikivie_0}. In this case, the magnetometer may be sensitive to magnetic fields as low as $\sim 10^{-18}$~T~\cite{magnetometer}.

For the sake of concreteness, let us consider a few options of existing magnets in which our set-up could be implemented. From here on we assume that the frequency of the external actuator can be made to approximately match the resonance frequency of the circuit, i.e., $\nu_r \approx \nu_{\rm act}$. The inductance of the pick-up loop is given by $L_m \simeq (\mu_0/\pi)\ell_m \log\left( r_m/ a_m \right)$, whereas the inductance of the small detection coil is $L_d = \mu_0 r_d c_d N_d^2$ with $c_d \simeq \log\left(8 r_d/ a_d \right) - 2$. Here $a_m$ and $a_d$ are the radii of the wires in the pick-up loop and detection coil, respectively. In the following we use $r_d = 0.5$~cm and $a_m = a_d = 1$~mm, so that, for $\ell_m \gg r_d$ and $\nu_r \sim 1$~Hz, we have $L_m + L_c \gg L_d$~\cite{Sikivie_0}.

\begin{table}
\begin{tabular}{|c|c|c|c|c|c|}
\hline
Magnet    & $B_{\rm ext}$(T)  & $\ell_m$(m)  & $r_m$(cm)  & $L_m$($\mu$H)  &  $k_{\rm AF}^{\rm 0} \left( {\rm GeV} \right)$  \\ \hline\hline
ADMX~\cite{ADMX} & 8 & 1 & 10  & 1.8 & $6.9 \times 10^{-35}$  \\ \hline
CMS~\cite{CMS}      &  4 & 13 & 10  & 24 & $3.4 \times 10^{-35}$ \\ \hline
NHMFL~\cite{wide_bore}  &  21 & $\sim 1$ & 5 & 1.6 &  $9.8 \times 10^{-35}$ \\
\hline
\end{tabular}
\caption{Basic parameters of the magnets considered in the text. In order to limit the necessary frequency of the varying side of the pick-up loop, we have restricted the ranges to $x_{\rm max} = 10$~cm for the ADMX and CMS magnets, and $x_{\rm max} = 5$~cm for the ultra wide-bore magnet at the NHMFL. We assumed that the coaxial cable connecting the pick-up loop to the detection circuit has an inductance $L_c \simeq 0.5 \; \mu$H~\cite{Sikivie_0}.}\label{tab}
\end{table}

Here we explictly consider the magnets from the ADMX~\cite{ADMX} and CMS~\cite{CMS} experiments, as well as the ultra wide-bore magnet at the NHMFL~\cite{wide_bore}. For further options, see ref.~\cite{Battesti}. In order to estimate the sensitivities, we assume that the respective bore volumes can be efficiently cooled down to $T = 0.4$~mK~\cite{BT} and that the superconducting circuit has $Q = 10^4$. Using ${\rm SNR} = 5$ as a threshold~\cite{Sikivie_0, Arias_0}, we find the sensitivities for $k_{\rm AF}^{\rm 0}$ listed in table~\ref{tab}, where the relevant parameters are summarized. Since $k_{\rm AF}^{\rm T}$ is in principle zero and $|{\bm \beta}| \simeq 10^{-4}$, cf. eqs.~\eqref{beta_x}-\eqref{beta_z}, the detection sensitivity to the spatial components is 
\begin{equation}
|{\bf k}_{\rm AF}^{\rm SCF}| \lesssim 10^{-31} \; {\rm GeV} \; ,
\end{equation}
which varies anually with $\Omega_\oplus \approx 0.2 \;\mu$Hz~\cite{Mewes}. This sensitivity is approximately eight orders of magnitude tighter than the currently best upper bounds from laboratory-based tests~\cite{Tables, Yuri}.

%\begin{equation}
%\delta I_{\rm T} = 2.96 \times 10^{-13} \; {\rm A} \; \sqrt{ \left( \frac{{\rm MHz}}{\omega_r} \right) \left( \frac{\mu{\rm H}}{L} \right) \left( \frac{Q}{10^4} \right) \left( \frac{T}{{\rm mK}} \right)    }
%\end{equation}

%%%%%%%%%%%%%%%%%%%%%%%%%%%%
\section{Concluding remarks} \label{sec_conc}
\indent

In this paper we analysed the consequences of the Carroll-Field-Jackiw model in the context of classical electrodynamics. In the presence of a strong external magnetic field, the CFJ 4-vector allows  the appearance of the displacement current ${\bf J}_{\rm CFJ}$, eq.~\eqref{CFJ_current}, which serves as source of an azimuthal LSV magnetic field ${\bf B}_{\rm CFJ}$, eq.~\eqref{B_CFJ}. Its flux through a carefully placed superconducting pick-up loop with a varying area would generate a current, which in turn induces a magnetic field in a small detection coil. This field could then be measured by a very sensitive magnetometer (e.g., a SQUID).

The proposed modification of the Sikivie-Sullivan-Tanner set-up increases the signal frequency by means of an external actuator allowing the area of the pick-up loop -- and the flux -- to oscillate harmonically at $\nu_r \sim 1$~Hz. The frequency of the LC circuit is made to match that of the varying area of the pick-up loop by means of a very large capacitance, thus allowing the signal to resonate, whereby the sensitivities can reach $10^{-31} \; {\rm GeV}$ for the magnets considered. It is worthwhile noting that, if we allow a non-zero time component in the SCF and neglect the anisotropic part of $k_{\rm AF}^0$, we find $k_{\rm AF}^{\rm T} \lesssim 10^{-35} \; {\rm GeV}$. This is ten orders of magnitude more sensitive than the results reported in ref.~\cite{Yuri}.

Solenoids with large bore volumes and high magnetic fields operating at sub-mK temperatures improve the attainable sensitivities, but the signal frequency is a crucial limiting factor for the proposed set-up. As shown in eq.~\eqref{SNR}, if everything else is held fixed, we have ${\rm SNR} \sim \sqrt{\nu_r / {\rm Hz}}$, which shows that the sensitivity scales only weakly with the frequency. This implies that a significant improvement in the sensitivity would require a much larger increase in the driving frequency of the external actuator. Nonetheless, it could pose problems with the mechanical integrity of the pick-up loop, as well as jeopardize the efficiency of the cooling due to friction.

A possible alternative that does not require sliding parts of the circuit would be to rotate the pick-up loop around $r_m/2$ at a few hundred rpm. However, for large $\ell_m$ this may become difficult due to centrifugal forces that may deform and potentially damage the wires. This can be minimized by restricting the dimensions of the circuit, but it would also reduce $V_m$.

Finally, we note that, even though the projected sensitivities represent a significant improvement over the currently best laboratory-based limits, they are weaker than the bounds from astrophysical and cosmological tests~\cite{Tables}. These use polarimetry data from distant sources to look for rotations in the polarization due to the birefringence induced by the modified dispersion relations~\cite{CFJ, Goldhaber, Kost5, Mewes2}.

The strength of astrophysical limits is not due to the precision of the underlying measurements, which is not particularly high~\cite{Kost5, WMAP}, specially in comparison with QED experiments. Instead, they exploit the dependence of the rotation $\Delta\chi$ on $r$, the large distance travelled from the sources to the observer via $\Delta\chi \simeq r \; k_{\rm AF} /2$~\cite{CFJ}. This means that these limits strongly lean on distance estimates over cosmological scales.

%Furthermore, polarimetry measurements depend on detailed knowledge of the intergalactic magnetic fields to control background effects, such as Faraday rotation.

 %the WMAP collaboration, for example, found a $\delta\psi \simeq 1.2^\circ \pm 2.2^\circ$ rotation in polarization.

Laboratory tests, on the other hand, generally rely on high-precision local measurements. The strongest bounds to date are extracted from hydrogen spectroscopy and measurements of polarization rotation in resonant cavities~\cite{Yuri}, where fractional uncertainties reach 1 part in $10^{15}$ and 1 part in $10^{10}$, respectively. The set-up envisaged here also owes its high projected sensitivity to exquisitely precise magnetometers capable of measuring fields as low as $\sim 10^{-18}$~T~\cite{magnetometer}. This performance level is only possible under exceptionally controlled conditions: well-understood backgrounds, high-precision mechatronics, and state-of-the-art cooling and shielding techniques, just to quote a few.

While not as sensitive as astrophysical limits, our set-up tests Lorentz invariance in a controlled environment, in which very precise measurements are feasible. Moreover, a laboratory experiment tests Lorentz invariance in our local neighborhood. Astrophysical tests, on the contrary, probe Lorentz violation over cosmological scales, where spacetime variations in the LSV backgrounds may be present. Hence, local tests like the one outlined here do play an important role in the search for new physics in our immediate vicinity.

In summary, though certainly challenging, small-scale dedicated laboratory-based experiments with cutting-edge design could provide the best terrestrial limits on the CFJ background, helping to cross-check the more indirect limits from astrophysical sources under controlled conditions.

%%%%%%%%%%%%%%%%%%%%%%%%%%%%%%%%%%%%%%

%Finally, we would like to point out that the LSV-modified Maxwell equations offer other measurement opportunities. Equation~\eqref{eq_max_amp} was used in a setting where only external magnetic fields are present, but the opposite situation could also be explored: if a very intense, time-varying external electric field ${\bf E}_{\rm ext}$ is produced in a source-free region, a second LSV displacement current ${\bf J}_{\rm CFJ}^{\prime} = -2{\bf k}_{\rm AF}\times{\bf E}_{\rm ext}$ would be induced. 

%Contrary to ${\bf J}_{\rm CFJ}$, which is aligned with ${\bf B}_{\rm ext}$, this novel current would generate a magnetic field with a non-trivial projection onto the fixed pick-up loop. The presence of ${\bf k}_{\rm AF}$ -- and not of its isotropic counterpart $k_{\rm AF}^0$ -- implies that the flux would vary with time also due to the instantaneous orientation of the background relative to the apparatus. This would further reduce the detectable signal, but nonetheless provide a means to measure the spatial components of $k_{\rm AF}$ without the strong suppression due to Earth's boost factor relative to the SCF. A more detailed analysis is currently being undertaken and the results will be reported elsewhere.

%%%%%%%%%%%%%%%%%%%%%%%%%%%%%%
\begin{acknowledgments}
The author thanks J.A. Helay\"el-Neto, M.M. Candido, J.T. Guaitolini Jr., G.P. de Brito, C.A. Zarro, D. Kroff, J. Jaeckel and G. Rikken for interesting discussions, as well as the anonymous referees for the constructive criticism. He is also grateful to COSMO - CBPF for the hospitality.
\end{acknowledgments}

%%%%%%%%%%%%%%%%%%%%%%%%%%%%%%%%%%%%%%%%%%%%
%\appendix

%%%%%%%%%%%%%%%%%%%%%%
%\section{Details} \label{app_1}
%\indent 

%%%%%%%%%%%%%%%%%%%%%%%%%%%%%%%%%%%%%%%%%%%%%%%%%

\end{document}